\documentclass[aps,prb,twocolumn,superscriptaddress]{revtex4}
\usepackage{graphicx}
\begin{document}
\title{Maximum supercurrent in Josephson junctions with alternating
critical current density}
\author{Maayan Moshe}
\affiliation{School of Physics and Astronomy, Raymond and Beverly
Sackler Faculty of Exact Sciences,\\ Tel Aviv University, Tel Aviv
69978, Israel}
\author{C. W. Schneider}
\affiliation{Experimentalphysik VI, Center for Electronic
Correlations and Magnetism, Institute of Physics,\\ Augsburg
University, D-86135 Augsburg, Germany}
\author{G. Bensky}
\affiliation{School of Physics and Astronomy, Raymond and Beverly
Sackler Faculty of Exact Sciences,\\ Tel Aviv University, Tel Aviv
69978, Israel}
\author{R. G. Mints}
\email[]{mints@post.tau.ac.il}
\affiliation{School of Physics and Astronomy, Raymond and Beverly
Sackler Faculty of Exact Sciences,\\ Tel Aviv University, Tel Aviv
69978, Israel}
\par
\date{\today}
\begin{abstract}
We consider theoretically and numerically magnetic field dependencies
of the maximum supercurrent across Josephson tunnel junctions with
spatially alternating critical current density. We find that two
flux-penetration fields and one-splinter-vortex equilibrium state exist
in long junctions.
\end{abstract}
\maketitle
\section{Introduction}
\label{sec_I}
\par
\begin{figure}
\includegraphics[width=0.60\columnwidth]{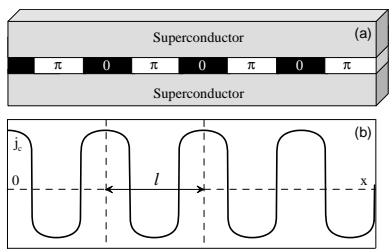}
\caption{Schematic drawings of: (a) sequence of $0\,$- and $\pi\,$- biased
Josephson junctions; (b) critical current density, $j_c(x)$.}
\label{fig_1}
\end{figure}

Studies of periodic or almost periodic Josephson tunnel structures
arranged in sequences of interchanging $0\,$- and $\pi\,$- biased
Josephson junctions (as shown in Fig. \ref{fig_1}) recently became a
subject of growing interest. These complex Josephson systems are
intensively treated experimentally, theoretically, and numerically in:
(a) superconductor-ferromagnet-superconductor (SFS) junctions in thin
films\cite{Bulaevskii_1, Buzdin_1, Ryazanov_1, Kontos_1, Blum_1} and
(b) Josephson grain boundaries in thin films of high-temperature
cooper-oxide superconductor YBa$_2$Cu$_3$O$_{7-x}$.\cite{Copetti_1,
Hilgenkamp_1, Harlingen_1, Tsuei_1, Hilgenkamp_2, Mannhart_1, Mints_1,
Mints_2, Mints_3, Buzdin_2}
\par
%
Equilibrium states of SFS Josephson junctions with a $\pi$-shift in the
phase difference between the superconducting banks has been predicted
almost three decades ago.\cite{Bulaevskii_1, Buzdin_1} However, only
recently SFS $\pi\,$- shifted junctions and SFS heterostructures of
interchanging $0\,$- and $\pi\,$- shifted fragments were studied
experimentally for the first time. \cite{Ryazanov_1, Kontos_1, Blum_1}
\par
The studies of Josephson properties of the asymmetric grain boundaries
in  YBa$_2$Cu$_3$O$_{7-x}$ thin films reveal an interesting and
important example of a Josephson system being an interchanging sequence
of $0\,$-$\pi$ biased junctions.\cite{Copetti_1, Hilgenkamp_1,
Harlingen_1, Tsuei_1, Hilgenkamp_2} The structure of these boundaries
is created by facets with a variety of orientations and lengths $l\sim
10 - 100\,$nm.\cite{Hilgenkamp_2} This grain boundary structure in
conjunction with the $d_{x^2-y^2}$-wave symmetry of the order
parameter\cite{Harlingen_1, Tsuei_1} can be considered as a Josephson
tunnel junction with spatially alternating critical current density,
$j_c(x)$, where $x$ axis is along the grain boundary.\cite{Harlingen_1,
Tsuei_1, Hilgenkamp_2} These rapid alternations with a typical
length-scale of $l$ significantly suppress the maximum supercurrent
$I_{\rm m}$ across the grain boundaries. This suppression is most
effective for the asymmetric 45$^{\circ}$ [001]-tilt grain boundaries
in YBa$_2$Cu$_3$O$_{7-x}$ films.\cite{Hilgenkamp_1, Hilgenkamp_2}
\par
The asymmetric 45$^{\circ}$ [001]-tilt grain boundaries in thin
YBa$_2$Cu$_3$O$_{7-x}$ films exhibit several remarkable and important
anomalies. First, the dependence of the maximum supercurrent $I_{\rm
m}$ on the applied magnetic field $H_a$ is
non-Fraunhofer.\cite{Copetti_1, Hilgenkamp_1, Mints_1, Buzdin_2}
Contrary to the classical Fraunhofer pattern with the central major
peak two symmetric major side-peaks appear at the two fields
$\pm\,H_{\rm sp}\ne 0$. Second, a spontaneous rapidly alternating
magnetic flux is generated at the grain boundaries.\cite{Mannhart_1}
Third, unquantized spontaneous flux structures include fragments formed
by pairs of single Josephson-type vortices carrying fluxes
$\phi_1<\phi_0/2$ and $\phi_2>\phi_0/2$. \cite{Mints_2, Mints_3} These
fluxes are complimentary and sum to $\phi_0$, {\it i.e.}, $\phi_1
+\phi_2 =\phi_0$ and therefore introduce splintered Josephson vortices.
It is worth noting here that the anomalous patterns $I_m(H_a)$ and the
unquantized splinter vortices appear under conditions of existence of
equilibrium spontaneous flux.
\par
In many cases the length-scale $l$ of the spatial alternations of the
critical current density $j_c(x)$ is bigger or much bigger than the
London penetration depth, $\lambda$, and is smaller or much smaller
than the local Josephson penetration depth, $\lambda_J$, defined by the
average of the {\it absolute} value of the critical current density. In
the limit of $l\ll\lambda_J$ the phase difference between the banks of
the tunnel junction, $\varphi (x)$, can be written as a sum of smooth,
$\psi (x)$, and rapidly varying, $\xi (x)$, terms.\cite{Mints_2}
Coarse-graining the phase $\varphi (x)$ over a distance ${\cal L}\gg l$
allows to consider the two terms $\psi (x)$ and $\xi (x)$ separately
from each other in the inner part of the junction. In this
approximation the coupling of $\psi (x)$ and $\xi (x)$ happens because
of the boundary conditions at the edges of the junction.
\par
In this paper we calculate both theoretically and numerically the
anomalous magnetic field dependence of the maximum supercurrent, $I_m$,
in Josephson tunnel junctions with spatially alternating critical
current density. The applied magnetic field $H_a$ is supposed to be
lower than the side-peaks field, {\it i.e.}, $|H_a|\ll H_{\rm
sp}\sim\phi_0/2\pi\lambda\, l$.
\par
The paper is organized as follows. In Sec. II we discuss the
coarse-grained equations for the phase difference across the banks of
Josephson junctions with alternating critical current density and
derive the boundary conditions to these equations. In Sec. III we
consider the maximum supercurrent across Josephson junctions
theoretically in two limiting cases of short and long junctions in low
and high magnetic fields. In Sec. IV we report on the results of
numerical simulations of the maximum supercurrent dependence on the
applied magnetic field. Sec. VII summarizes the overall conclusions.
\par
\section{Coarse-grained equations}
\label{sec_II}
\par
We treat a one-dimensional Josephson junction parallel to the $x$ axis
with the tunneling current density ${\bf j}\,\|\,\hat{\bf y}$,
$j_y(x)=j(x)$, and the magnetic field ${\bf H}\,\|\,\hat{\bf z}$,
$H_z(x)=H(x)$. Assume also that the critical current density $j_c(x)$
is an alternating periodic or almost periodic function taking positive
and negative values with a typical length-scale $l$. The geometry of
the problem is shown schematically in Fig. \ref{fig_1}.
\par
First, we introduce the average value of the critical current density,
$\langle\, j_c\rangle$, the effective Josephson penetration depth,
$\Lambda$, defined by $\langle\, j_c\rangle$, and the local Josephson
penetration depth, $\lambda_J$, defined by the average value of $|j_c|$
\begin{eqnarray}
\label{eqn_01}
\langle\, f\rangle &=& {1\over L}\int_0^L f(x)\,dx\,,\\
\label{eqn_02}
\Lambda &=& \sqrt{c\phi_0\over 16\pi^2\lambda\,\langle\, j_c\rangle}\,,\\
\label{eqn_03}
\lambda_J &=& \sqrt{c\phi_0\over 16\pi^2\lambda\langle\,|j_c|\rangle}\,,
\end{eqnarray}
where Eq. (\ref{eqn_01}) is the definition of averaging, $L$ is the
length of the junction ($L\gg l$), $\phi_0$ is the flux quantum, and
$\lambda$ is the London penetration depth.
\par
Next, we assume that $\lambda\ll l\ll\lambda_J\ll\Lambda$. In this case
the phase difference $\varphi (x)$ satisfies the equation
\begin{equation}
\label{eqn_04}
\Lambda^2\varphi ''-{j_c(x)\over\langle j_c\rangle}\,\sin\varphi=0.
\end{equation}
It is convenient for the following analyses to write the critical
current density $j_c(x)$ in the form
\begin{equation}
\label{eqn_05}
j_c(x)=\langle j_c\rangle\,[1+g(x)]
\end{equation}
introducing a rapidly alternating function $g(x)$ with a zero average
value, $\langle g(x)\rangle=0$, and a typical length-scale of order
$l$. It is worth noting that $g(x)$ is a unique internal characteristic
of a junction. Using the function $g(x)$ we rewrite Eq. (\ref{eqn_04})
as
\begin{equation}
\label{eqn_06}
\Lambda^2\varphi ''-[1+g(x)]\,\sin\varphi=0\,.
\end{equation}
\par
The idea of the following calculation is based on a mechanical analogy
(Kapitza's pendulum).\cite{Landau_1,Arnold_1} Two types of terms appear
in Eq. (\ref{eqn_06}): fast terms alternating over a length $l$ and
smooth terms varying over a length $\Lambda\gg l$. The fast alternating
terms cancel each other, independently of the smooth terms, which also
cancel each other.
\par
Thus, to find solutions of Eq. (\ref{eqn_06}) we use the ansatz
\begin{equation}
\label{eqn_07}
\varphi (x)=\psi (x)+\xi (x)\,,
\end{equation}
where $\psi (x)$ is a smooth function with the length-scale of order
$\Lambda$, $\xi (x)$ is a rapidly alternating function with the
length-scale of order $l$, and the variations of $\xi (x)$ are small,
{\it i.e.}, $\langle|\xi (x)|\rangle\ll 1$.\cite{Mints_2} We assume
also that the average value of $\xi (x)$ is zero, $\langle\xi
(x)\rangle =0$. It is worth mentioning that the ansatz given by Eq.
(\ref{eqn_07}) is similar to the one used to solve the Kapitza's
pendulum.\cite{Landau_1, Arnold_1}
\par
Substituting Eq.~(\ref{eqn_07}) into Eq.~(\ref{eqn_06})
and keeping terms up to first order in $\xi (x)$ we find \cite{Mints_2}
\begin{eqnarray}
\label{eqn_08}
\Lambda^2\psi'' - {j_\psi (x)\over\langle j_c\rangle} &=& 0\,,\\
\label{eqn_09}
\Lambda^2\xi'' - {j_\xi (x)\over\langle j_c\rangle} &=& 0\,,
\end{eqnarray}
where the smooth $j_\psi (x)$ and alternating $j_\xi (x)$ components
of the tunneling current density $j= j_\psi +j_\xi$ are
\begin{eqnarray}
\label{eqn_10}
j_\psi &=& \langle j_c\rangle \left(\sin\psi -
\gamma\sin\psi\cos\psi\right)\,,\\
\label{eqn_11}
j_\xi &=& \langle j_c\rangle\,g(x)\sin\psi\,,
\end{eqnarray}
the dimensionless constant $\gamma$ is equal to
\begin{equation}
\label{eqn_12}
\gamma = \langle g(x)\xi_g (x)\rangle\,,
\end{equation}
and the rapidly alternating phase $\xi_g(x)$ is defined by
\begin{equation}
\label{eqn_13}
\xi (x) = -\xi_g (x)\sin\psi\,.
\end{equation}
It follows from Eqs. (\ref{eqn_09}), (\ref{eqn_11}) and (\ref{eqn_13})
that
\begin{equation}
\label{eqn_14}
\Lambda^2\xi''_g + g(x) = 0\,,
\end{equation}
{\it i.e.}, the rapidly alternating phase shift $\xi_g$ depends only on
the effective penetration depth $\Lambda$ and the function $g(x)$.
Therefore, the phase $\xi_g(x)$ is an internal characteristics of a
junction.
\par
It follows from Eq. (\ref{eqn_10}) that the smooth current density
$j_\psi$ includes the initial first harmonic term $\propto\sin\psi$ and
an additional second harmonic term $\propto\sin 2\psi$, which results
from constructive interference of the rapidly alternating critical
current density $\propto g(x)$ and phase $\xi (x)$.
\cite{Mints_2}
\par
To summarize the derivation of the system of coarse-grained equations
(\ref{eqn_08})--(\ref{eqn_11}) it is worth noting that the typical
value of $\xi_g(x)$ is small, but at the same time the typical value of
$g(x)$ is big, {\it i.e.}, $\langle|\xi_g(x)|\rangle\ll 1$ and
$\langle|g(x)|\rangle\gg 1$. As a result, the dimensionless parameter
$\gamma$, which is proportional to the average of the product of the
two rapidly alternating functions $\xi_g(x)$ and $g(x)$ might be of the
order of unity.\cite{Mints_2,Mints_3}
\par
The energy ${\cal E}$ of a junction with alternating critical current
density $j_c(x)$ yields
\begin{equation}
\label{eqn_15}
{\cal E}={\hbar\langle j_c\rangle\over 2e}
\int_0^L \!\!\left({\Lambda^2\over 2}\,\psi'^2 + 1-\cos\psi -
{\gamma\over 2}\sin^2\psi\right)dx.
\end{equation}
The last term in the integral in Eq. (\ref{eqn_15}) is for the
contribution of both the fast alternating current $j_\xi(x)$ and phase
$\xi(x)$. It is worth noting that minimization of the functional ${\cal
E}\{\psi\}$ results in Eq. (\ref{eqn_08}) for the phase $\psi(x)$.
\par
It follows from Eqs. (\ref{eqn_08}) and (\ref{eqn_15}) that if the
parameter $\gamma
>1$, then there are two series of stable uniform equilibrium states with
$\psi_e =2\pi n \pm\psi_\gamma$ and current density $j_\psi(\psi_e)=0$,
where $n=0,\pm1,\pm 2,\,\dots$ is an integer and the phase
$\psi_\gamma$ is defined by\cite{Mints_2}
\begin{equation}
\label{eqn_16}
\gamma\cos\psi_\gamma =1\,.
\end{equation}
All equilibrium states with $\psi =\psi_e$ have the same energy
\begin{equation}
\label{eqn_17}
{\cal E}_\gamma =-{\hbar\langle j_c\rangle\over 2e}\,{(\gamma -1)^2
\over 2\gamma}\,L\,,
\end{equation}
which is less then the energy ${\cal E}_0=0$ of the series of unstable
states with the phase $\psi =2\pi n$.\cite{Mints_2} If the parameter
$\gamma<1$, then there is only one series of stable uniform equilibrium
states with $\psi_0 =2\pi n$ and ${\cal E}_0=0$.
\par
The two series of stable equilibrium states result in existence of two
different single Josephson vortices (two {\it
splinters}).\cite{Mints_2,Mints_3} The phase $\psi(x)$ for the first
(``small'') splinter vortex varies from $-\psi_\gamma$ at $x=-\infty$
to $\psi_\gamma$ at $x=+\infty$. This vortex carries flux $\phi_1
=\phi_0\psi_\gamma/\pi\le\phi_0/2$. The phase for the second (``big'')
splinter vortex varies from $\psi_\gamma$ at $x=-\infty$ to $2\pi
-\psi_\gamma$ at $x=\infty$. This vortex caries flux $\phi_2
=\phi_0(\pi -\psi_\gamma)/\pi\geq \phi_0/2$. As a result any flux
structure inside a junction with an alternating critical current
density and with $\gamma > 1$ consists of series of interchanging small
and big splinter vortices. \cite{Mints_2,Mints_3} It is also important
mentioning that $\phi_1 +\phi_2 =\phi_0$.
\par
Consider now the boundary conditions to Eq. (\ref{eqn_08}), {\it i.e.},
for the smooth phase shift $\psi (x)$. Using equations
\begin{eqnarray}
\label{eqn_18}
H&=&{\phi_0\over 4\pi\lambda}\,{d\varphi\over dx}\,,\\
\label{eqn_19}
\varphi(x) &=& \psi(x) - \xi_g(x)\,\sin\psi(x)
\end{eqnarray}
we find the boundary conditions for $\psi (x)$ in the form
\begin{eqnarray}
\label{eqn_20}
\varphi'(0)&=&\psi'_0 - \xi'_{\rm g0}\sin\psi_0 =
{4\pi\lambda\over\phi_0}\,H_0\,, \\
\label{eqn_21}
\varphi'(L)&=&\psi'_L - \xi'_{\rm gL}\sin\psi_L =
{4\pi\lambda\over\phi_0}\,H_L\,,
\end{eqnarray}
where $\psi_0=\psi(0)$, $\psi_L=\psi(L)$, $\psi'_0=\psi'(0)$,
$\psi'_L=\psi'(L)$, $H_0=H(0)$, $H_L=H(L)$, $\xi'_{\rm g0}=\xi_g'(0)$,
and $\xi'_{\rm gL}=\xi'_g(L)$. Next, we use the fact that the average
value of $g(x)$ is zero and integrate Eq. (\ref{eqn_14}) from $0$ to
$L$. This leads to
\begin{equation}
\label{eqn_22}
\xi'_{\rm g0} = \xi'_{\rm gL}=\xi'_{\rm gb},
\end{equation}
where $\xi'_{\rm gb}$ is an internal parameter characterizing the edges
of the junction. Now the boundary conditions given by Eqs.
(\ref{eqn_20}) and (\ref{eqn_21}) take the form
\begin{eqnarray}
\label{eqn_23}
\psi'_0 - \xi'_{\rm gb}\sin\psi_0 &=& {4\pi\lambda\over\phi_0}\,H_0\,,\\
\label{eqn_24}
\psi'_L - \xi'_{\rm gb}\sin\psi_L &=&{4\pi\lambda\over\Phi_0}\,
H_L\,.
\end{eqnarray}
\par
Compare now the values of the derivatives $\psi_0'$, $\psi'_L$, and
$\xi'_{\rm gb}$. Using Eqs. (\ref{eqn_09}), (\ref{eqn_12}) and
(\ref{eqn_14}) we obtain
\begin{equation}
\label{eqn_25}
\gamma = \langle g(x)\,\xi_g \rangle = -\langle \Lambda^2 \xi''_g
(x)\xi_g(x)\rangle =
 \langle\left[\Lambda\xi_g' (x)\right]^2\rangle
\end{equation}
and arrive to the relation
\begin{equation}
\label{eqn_26}
\Lambda\,|\xi_g'(x)|\sim\sqrt{\gamma}\sim 1.
\end{equation}
A similar estimate $\Lambda |\psi'(x)|\sim 1$ follows from Eqs.
(\ref{eqn_08}) and (\ref{eqn_10}). These estimates demonstrate that
both derivatives $\psi'(x)$ and $\xi'_g(x)$ are of the same order
although $\langle|\xi_g(x)|\rangle\ll\langle|\psi(x)|\rangle$. Indeed,
for a typical junction exhibiting spontaneous equilibrium flux we have
$\gamma\sim 1$.\cite{Mints_3}
\par
The fact that $\Lambda\xi'_{\rm gb}\sim 1$ makes it convenient for the
following analysis to write the derivative $\xi'_{\rm gb}$ in the form
\begin{equation}
\label{eqn_27}
\xi'_{\rm gb} ={\alpha\over\Lambda},
\end{equation}
where $\alpha\sim 1$ is an internal parameter characterizing the edges
of the junction.
\par
Thus, in the framework of the coarse-grained approach a junction with
an alternating critical current density is characterized by two
dimensionless parameters $\alpha$ and $\gamma$.
\par
Assume, that the current across a junction $I\ne 0$, then we have the
relations
\begin{eqnarray}
\label{eqn_28}
H_0=H_a+{2\pi\over c}I\,,\\
\label{eqn_29}
H_L=H_a-{2\pi\over c}I.
\end{eqnarray}
In this case the boundary conditions given by Eqs. (\ref{eqn_23}) and
(\ref{eqn_24}) take the final form
\begin{eqnarray}
\label{eqn_30}
\psi'_0={4\pi\lambda\over\phi_0}\,H_a + {8\pi^2\lambda\over
c\phi_0}\,I + {\alpha\over\Lambda}\sin\psi_0\,,\\
\label{eqn_31}
\psi'_L ={4\pi\lambda\over\phi_0}\,H_a - {8\pi^2\lambda\over
c\phi_0}\,I + {\alpha\over\Lambda}\sin\psi_L\,.
\end{eqnarray}
\par
The fact that the rapidly alternating critical current density $j_c(x)$
has low average value $[\langle j_c(x)\rangle \ll
\langle|j_c(x)|\rangle]$ might significantly affect the maximum
supercurrent. Indeed, assume that the Josephson current density
includes both the first and the second harmonics,\cite{Golubov_1} {\it
i.e.},
\begin{equation}
\label{eqn_32}
j= j_{\rm c1}(x)\sin\varphi + j_{\rm c2}\sin 2\varphi\,,
\end{equation}
where $j_{\rm c1}(x)$ is rapidly alternating along the junction and
$j_{\rm c2}$ is spatially independent.
\par
In this case the coarse-graining approach remains the same as above.
The effect of the second harmonics on the maximum supercurrent $I_m$
increases with the increase of the dimensionless parameter
$\gamma_2=j_{\rm c2}/ \langle j_{\rm c1}\rangle$. The value of
$\gamma_2$ might be of order of unity and higher even if $j_{\rm c2}$
is low compared to $\langle |j_{\rm c1}(x)|\rangle$.
\par
\section{Maximum Supercurrent}
\label{sec_III}
The Josephson tunneling current, $I$, across Josephson tunnel junction
with an alternating critical current density can be written as a sum of
two terms $I_\psi$ and $I_\xi$
\begin{equation}
\label{eqn_33}
I=\int_0^L j\,dx = I_\psi + I_\xi\,,
\end{equation}
where the currents $I_\psi$ and $I_\xi$ are given by
\begin{eqnarray}
\label{eqn_34}
I_\psi &=& \int_0^L j_\psi\,dx = I_c\,\Lambda
\left(\psi'_L - \psi'_0\right)\,, \\
\label{eqn_35}
I_\xi &=& \int_0^L j_\xi\,dx =\alpha I_c\,(\sin\psi_0 -\sin\psi_L)\,,
\end{eqnarray}
and the current $I_c$ is defined as
\begin{equation}
\label{eqn_36}
I_c = \Lambda\langle j_c\rangle\,.
\end{equation}
It follows from Eqs. (\ref{eqn_34}) and (\ref{eqn_35}) that both
$I_\psi$ and $I_\xi$ are defined by the smooth phase $\psi$ only.
\par
Magnetic flux inside the junction
\begin{equation}
\label{eqn_37}
\phi =2\lambda\,\int_0^L H\,dx ={\phi_0\over 2\pi}\,(\psi_L -\psi_0)
\end{equation}
results in the phase difference
\begin{equation}
\label{eqn_38}
\psi_L -\psi_0 =2\pi\,{\phi\over\phi_0}\,.
\end{equation}
Using Eqs. (\ref{eqn_35}) and (\ref{eqn_38}) we obtain the current
$I_\xi$ as a function of the flux inside the junction
\begin{eqnarray}
\label{eqn_39}
I_\xi &=& \alpha I_c\bigg[ \sin\psi_0 -\sin\bigg(\psi_0
+2\pi{\phi\over\phi_0}\bigg) \bigg]=
\nonumber\\
&=& -2\alpha I_c\sin\bigg(\pi{\phi\over\phi_0}\bigg)
\cos\bigg(\psi_0+\pi{\phi\over\phi_0}\bigg)\,.
\end{eqnarray}
\par
In order to calculate the total current $I_\psi$ one has to know the
spatial distribution of the phase $\psi (x)$ in detail.
\par
In what follows we calculate the maximum supercurrent $I_m$
theoretically in the limiting cases of short $(L\ll\Lambda$) and long
($L\gg\Lambda$) junctions treating the problem separately for the
Meissner and mixed states.
\subsection{Maximum current across short junctions}
\label{sec_III_a}
We calculate now the maximum supercurrent $I_m(H_a)$ of a short
junction, $L\ll\Lambda$. In this case the spatial dependence of the
smooth phase $\psi (x)$ in the main approximation in $L/\Lambda\ll 1$
is linear
\begin{equation}
\label{eqn_40}
\psi (x)=\psi_0 + 2\pi\,{\phi_i\over\phi_0}\,{x\over L},
\end{equation}
where
\begin{equation}
\label{eqn_41}
\phi_i = 2\lambda LH_i
\end{equation}
is the ``internal'' flux and $H_i$ is the magnetic field inside the
junction. Next, we use Eqs. (\ref{eqn_30}), (\ref{eqn_31}),
(\ref{eqn_34}), (\ref{eqn_35}), and (\ref{eqn_40}) and obtain the
following relations
\begin{eqnarray}
\label{eqn_42}
H_i&=&H_a,\\
\label{eqn_43}
\psi_L&=&\psi_0 + 2\pi\,{\phi_a\over\phi_0}, \quad \phi_a=2\lambda LH_a\,,\\
\label{eqn_44}
I&=&\alpha I_c\,\left(\sin\psi_L - \sin\psi_0\right)\,.
\end{eqnarray}
Combining Eqs. (\ref{eqn_43}) and (\ref{eqn_44}) we find that the
maximum value of the total current $I(\phi_a)$ is given by
\begin{equation}
\label{eqn_45}
I_m=2\alpha I_c
\left|\,\sin\left(\pi{\phi_a\over\phi_0}\right)\right|\,.
\end{equation}
\par
It follows from Eq. (\ref{eqn_45}) that the maximum supercurrent across
short junctions with spatially alternating critical current density is
defined only by the surface current $I_\xi$ (in the main approximation
in $L/\Lambda\ll 1$). As a result the dependence $I_m(\phi_a)$ is
obviously non-Fraunhofer. The value of $I_m$ is oscillating
periodically in $\phi_a$ with the period that is equal to flux quantum
$\phi_0$. Contrary to the case of a constant critical current density
the amplitude of oscillations of $I_m$ is not decreasing with the
increase of the applied field $H_a$.\cite{Kulik_1, Barone_1}
\par
\subsection{Meissner and mixed states in long junctions}
\label{sec_III_b}
In this subsection we consider the spatial distributions of the phase
difference and the flux in long junctions, $L\gg\Lambda$. We start with
the low field limit, {\it i.e.}, we assume that the applied field $H_a
\ll H_s$, where
\begin{equation}
\label{eqn_46}
H_s = {\phi_0\over 2\pi \lambda\Lambda}
\end{equation}
is the flux penetration field for a long junction with a constant
critical current density $j_c ={\rm const}$.\cite{Kulik_1,Barone_1} In
the following analysis we use an approach similar to the one, which was
first developed by Owen and Scalapino. \cite{Owen_1}
\par
\begin{figure}
\includegraphics[width=0.60\columnwidth]{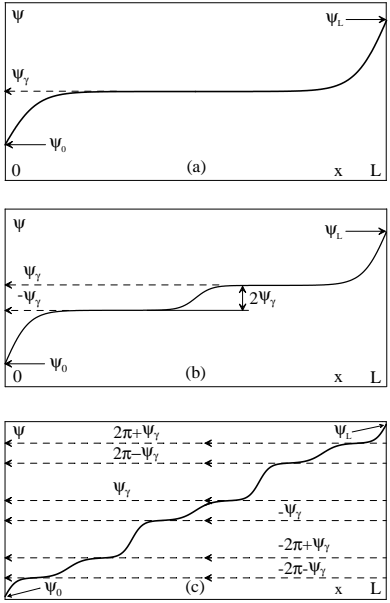}
\caption{Spatial distributions of the phase $\psi(x)$ in a long junction
for different values of the applied field $H_a$ and the internal flux
$\phi_i$. (a) $H_a < H_{\rm s1}$, $\phi_i =0$, (b) $H_{\rm s1}\le
H_a\le H_{\rm s2}$, $\phi_i =\phi_1$, (c) $H_a>H_{\rm s2}\,$,
$\phi_i\approx\phi_a$.}
\label{fig_2}
\end{figure}
%
In the case of $L\gg\Lambda$ and $H_a\ll H_s$ the total supercurrent
$I=I_\psi + I_\xi$ is a surface current localized in a layer with a
width $\sim l\gg\lambda$. It follows from Eqs. (\ref{eqn_34}) and
(\ref{eqn_35}) that in order to calculate $I_\psi$ and $I_\xi$ we have
to find the dependencies of $\psi'_L$ and $\psi'_0$ on $\psi_0$ and
$\psi_L$. These dependencies are given by the first integral of Eq.
(\ref{eqn_08})
\begin{equation}
\label{eqn_47}
{\Lambda^2\over 2}\psi'^2 + \cos\psi -{\gamma\over 4}\cos2\psi = {\rm
const}\,.
\end{equation}
It is worth mentioning that Eq. (\ref{eqn_47}) describes the density of
the energy ${\cal E}$ given by Eq. (\ref{eqn_15}).
\par
The spatial distribution of $\psi(x)$ depends on the magnetic
prehistory of the sample. We begin here for brevity with the case of a
junction in the Meissner state. In this case the flux is localized at
the edges of the junction. As a result in a long junction the phase
$\psi (x)$ in the inner part equals to a certain constant
$\psi_\infty$. The first correction to this constant is proportional to
$\exp{(-L/\Lambda)}\ll 1$. In other words we have
\begin{equation}
\label{eqn_48}
\psi (L/2) =\psi_\infty\,, \quad \psi'(L/2)=0\,,
\end{equation}
where the phase $\psi_\infty$ is given by one of the stable equilibrium
values of $\psi$, {\it i.e.}, $\cos\psi_\infty =1/\gamma$. Combining
Eqs. (\ref{eqn_47}), (\ref{eqn_48}) and (\ref{eqn_16}) we find that the
constant in the RHS of Eq. (\ref{eqn_47}) is given by
\begin{equation}
\label{eqn_49}
{\rm const} =\cos\psi_\infty - {\gamma\over 4}\cos 2\psi_\infty =
{\gamma\over 4} + {1\over 2\gamma}\,.
\end{equation}
The above relation allows for transforming Eq. (\ref{eqn_47}) into
\begin{equation}
\label{eqn_50}
\Lambda^2\psi'^2 = \gamma \left(\cos\psi_\gamma -\cos\psi\right)^2.
\end{equation}
\par
We calculate first the flux penetration field into a junction with an
alternating critical current density and a {\it zero} total current,
$I=0$. The dependence $\psi (x)$ for this case is shown schematically
in Fig.\,\ref{fig_2}\,(a). It follows then from Eq. (\ref{eqn_50}) that
\begin{eqnarray}
\label{eqn_51}
\Lambda\psi'_0 =\sqrt{\gamma}(\cos\psi_0 - \cos\psi_\gamma)\,,\\
\label{eqn_52}
\Lambda\psi'_L =\sqrt{\gamma}(\cos\psi_\gamma - \cos\psi_L)\,.
\end{eqnarray}
Next, we combine Eqs. (\ref{eqn_30}), (\ref{eqn_51}), (\ref{eqn_31}),
and (\ref{eqn_52}) and obtain two relations between the applied field
$H_a$ and the phases $\psi_0$ and $\psi_L$
\begin{eqnarray}
\label{eqn_53}
H_a={H_s\over 2}\,\left[\sqrt{\gamma +\alpha^2}\,\cos\left(\psi_0 +
\psi_\alpha\right) - {1\over\sqrt{\gamma}}\right]\,,\\
\label{eqn_54}
H_a={H_s\over 2}\,\left[{1\over\sqrt{\gamma}}-\sqrt{\gamma +\alpha^2}\,
\cos\left(\psi_L - \psi_\alpha\right)\right]\,.
\end{eqnarray}
where we introduce the phase $\psi_\alpha$ as
\begin{equation}
\label{eqn_55}
\tan\psi_\alpha = {\alpha\over\sqrt{\gamma}}\,.
\end{equation}
\par
In the following analysis we assume, for definiteness, that
$\psi_\alpha < \psi_\gamma$. In this case the dependence $\psi(x)$
looks as shown schematically in Fig.~\ref{fig_2}.
\par
As a function of $\psi_0$ the RHS of Eq. (\ref{eqn_53}) is bounded. The
maximum field
\begin{equation}
\label{eqn_56}
H_{\rm s1}={H_s\over 2}\,\left[\sqrt{\gamma +\alpha^2}- {1\over
\sqrt{\gamma}}\,\right]
\end{equation}
is achieved at
\begin{equation}
\label{eqn_57}
\psi_0 = -\psi_\alpha\,.
\end{equation}
\par
Therefore, if the applied field $H_a$ reaches the value of $H_{\rm s1}$
then the Meissner state in a long junction becomes unstable and the
small splinter vortex\cite{Mints_2,Mints_3} carrying flux
\begin{equation}
\label{eqn_58}
\phi_1=\phi_0{\psi_\gamma\over\pi}\le\phi_0/2
\end{equation}
enters into the inner part of the junction as shown in
Fig.\,\ref{fig_2}\,(b). This feature is a direct consequence of
existence of the splinter vortices in junctions with $\gamma\ge 1$.
\par
It follows from Eq. (\ref{eqn_50}) that in this one-vortex state
\begin{equation}
\label{eqn_59}
\Lambda\psi'_{\rm 0,L} =\sqrt{\gamma}\left(\cos\psi_\gamma
-\cos\psi_{\rm 0,L}\right) >0\,.
\end{equation}
Using Eqs. (\ref{eqn_30}), (\ref{eqn_31}), and (\ref{eqn_59}) we obtain
the relations between the applied field $H_a$ and the phases $\psi_0$
and $\psi_L$
\begin{eqnarray}
\label{eqn_60}
H_a={H_s\over 2}\,
\left[{1\over\sqrt{\gamma}} -\sqrt{\gamma +\alpha^2}\,
\cos\left(\psi_0 -\psi_\alpha\right)\right]\,,\\
\label{eqn_61}
H_a={H_s\over 2}\,
\left[{1\over\sqrt{\gamma}} -\sqrt{\gamma +\alpha^2}\,
\cos\left(\psi_L -\psi_\alpha\right)\right]\,.
\end{eqnarray}
The RHS's of Eqs. (\ref{eqn_60}) and (\ref{eqn_61}) are bounded as
functions of $\psi_0$ and $\psi_L$ and the maximum field
\begin{equation}
\label{eqn_62}
H_{\rm s2}={H_s\over 2}\,
\left[\sqrt{\gamma +\alpha^2}+ {1\over\sqrt{\gamma}}\,\right]
\end{equation}
is achieved at $\psi_0 = \psi_\alpha -\pi +2\pi n$ and $\psi_L =
\psi_\alpha +\pi +2\pi m$  where $n, m =0,\pm 1,
\pm 2, ...$ are integers. If the applied field $H_a$ reaches the
value of $H_{\rm s2}$ the one-vortex state becomes unstable and
magnetic flux penetrates into the bulk until a mixed state with a
finite density of vortices is established (see Fig.~\ref{fig_2}\,(c)).
\par
Therefore, the rapid spatial alternations of the critical current
density $j_c(x)$ in case of $\gamma\ge 1$ lead to existence of a
specific equilibrium one-splinter-vortex state. This state appear if
the applied field $H_a$ is from the interval $H_{\rm s1}\le H_{\rm
a}\le H_{\rm s2}$. It is worth noting here that the case of a standard
Josephson junction ($j_c={\rm const}$) corresponds to $\alpha =0$ and
$\gamma =1$. It follows then from Eqs. (\ref{eqn_58}), (\ref{eqn_56}),
(\ref{eqn_62}), and (\ref{eqn_46}) that for these values of the
parameters $\alpha$ and $\beta$ we have $\phi_1 =0$, $\phi_2 =\phi_0$,
$H_{\rm s1}=0$, and $H_{\rm s2}=H_s$, {\it i.e.}, there is only one
Josephson vortex and the Meissner state exists if $0\le H_a\le H_s$ as
it has to be.\cite{Kulik_1} This verification means that the above
results are self-consistent in describing the case of a standard
Josephson junction.
\subsection{Maximum supercurrent in the Meissner state}
\label{sec_III_c}
We calculate now the maximum supercurrent $I_m$ in the Meissner state
in a long junction, {\it i.e.}, we assume that $L\gg\Lambda$ and the
smooth phase $\psi$ inside the junction is given by one of its
equilibrium values $\psi_e =2\pi n\pm\psi_\gamma$, where $n=0,\pm 1,\pm
2, ...$ is an integer. The spatial distribution of $\psi(x)$
corresponding to the current $I_m$ is shown in Fig. \ref{fig_2}\,(a).
It follows then from Eq. (\ref{eqn_50}) that
\begin{eqnarray}
\label{eqn_63}
\psi'_0 ={\sqrt{\gamma}\over\Lambda}\,(\cos\psi_\gamma -\cos\psi_0),\\
\label{eqn_64}
\psi'_L ={\sqrt{\gamma}\over\Lambda}\,(\cos\psi_L -\cos\psi_\gamma)\,.
\end{eqnarray}
Using Eqs. (\ref{eqn_63}), (\ref{eqn_64}) and the boundary conditions
given by Eqs. (\ref{eqn_30}) and (\ref{eqn_31}) we obtain equations
relating the current $I$, the applied field $H_a$ and the phases
$\psi_0$ and $\psi_L$
\begin{equation}
\label{eqn_65}
H_a ={1\over 2}\, H_m\left[\cos(\psi_L+\psi_\alpha)
-\cos(\psi_0-\psi_\alpha)\right]\,,
\end{equation}
\begin{equation}
\label{eqn_66}
{I\over I_c}={2\over\sqrt{\gamma}} -2\,{H_m\over H_s}
\left[\cos(\psi_0 -\psi_\alpha) +\cos(\psi_L +\psi_\alpha)\right]\,,
\end{equation}
where we introduce the field $H_m$ as
\begin{equation}
\label{eqn_67}
H_m ={{H_{\rm s1} +H_{\rm s2}}\over 2} = {H_s\over 2}\,\sqrt{\gamma
+\alpha^2}\,.
\end{equation}
\par
The two relations given by Eqs. (\ref{eqn_65}) and (\ref{eqn_66}) allow
to obtain the dependence of the current $I$ on the field $H_a$ and the
phase $\psi_0$ in the form
\begin{equation}
\label{eqn_68}
{I\over I_c}={2\over\sqrt{\gamma}} -4{H_a\over H_s} -4\,{H_m\over
H_s}\,\cos(\psi_0-\psi_\alpha)\,.
\end{equation}
It follows from Eq. (\ref{eqn_68}) that the maximum current $I_m$
corresponds to $\cos(\psi_0 -\psi_\alpha) =-1$. Combining the above
calculation valid for $H_a>0$ with the one valid for $H_a<0$ we obtain
the dependence $I_m(H_a)$ in its final form
\begin{equation}
\label{eqn_69}
I_m =4I_c\,{{H_{\rm s2} - |H_a|}\over H_s} ={c\over 2\pi}\,(H_{\rm s2}
- |H_a|)\,.
\end{equation}
\par
Thus, in the Meissner state the maximum value of $I_m$ is achieved at
$H_a =0$ and is equal to
\begin{equation}
\label{eqn_70}
I_m(0) ={cH_{\rm s2}\over 2\pi}= 2I_c\,\left[\sqrt{\gamma
+\alpha^2}+{1\over\sqrt{\gamma}}\right].
\end{equation}
\par
It is worth noting that for a standard Josephson junction ($\alpha =0,
\gamma =1$) and therefore we have $H_{\rm s2}=H_s$. As a result
Eqs. (\ref{eqn_69}) and (\ref{eqn_70}) coincide with the similar
equations that were first derived by Owen and Scalapino.\cite{Owen_1}
\subsection{Maximum supercurrent in the mixed state}
\label{sec_III_d}
We calculate now the maximum supercurrent, $I_m$, in long junctions
($L\gg\Lambda$) in the mixed state, {\it i.e.}, we assume that the
applied magnetic field $H_a$ is higher than $H_{\rm s2}$. In the mixed
state the field inside the junction, $H_i$, is almost uniform and $\psi
(x)$ takes the form
\begin{equation}
\label{eqn_71}
\psi=\psi_0+2{H_i\over H_s}{x\over\Lambda}\,.
\end{equation}
The dependence of the supercurrent on the applied field follows from
the boundary conditions (\ref{eqn_30}) and (\ref{eqn_31}) yielding the
system of equations
\begin{eqnarray}
\label{eqn_72}
\pi{\phi_a\over\phi_0}&=&\pi{\phi_i\over\phi_0}-{\alpha\over 2}
{L\over\Lambda}\sin\psi_m\cos\left(\pi{\phi_i\over\phi_0}\right),
\\
\label{eqn_73}
I&=&-2\alpha I_c\cos\psi_m\sin\left(\pi{\phi_i\over\phi_0}\right),
\end{eqnarray}
where the phase $\psi_m$ is defined as
\begin{equation}
\label{eqn_74}
\psi_m={\psi_0+\psi_L\over 2}\,.
\end{equation}
\par
Next, we use the Lagrange multipliers method to find the maximum of the
supercurrent defined by Eq. (\ref{eqn_73}) under the constraint given
by Eq. (\ref{eqn_72}) and arrive to
\begin{eqnarray}
\label{eqn_75}
{\phi_0\over\pi I_c }\,{\partial I\over\partial\phi_i} &=&{\cal
L}\,{\partial\phi_a \over\partial\phi_i}\,,
\\
\label{eqn_76}
{\phi_0\over\pi I_c}\,{\partial I\over\partial\psi_m} &=&{\cal
L}\,{\partial\phi_a\over\partial\psi_m}\,,
\end{eqnarray}
where ${\cal L}$ is the Lagrange multiplier to be determined. In the
main approximation in $\Lambda/L\ll 1$ the solution of Eqs.
(\ref{eqn_75}) and (\ref{eqn_76}) is given by
\begin{equation}
\label{eqn_77}
\cos\psi_m\cos\left(\pi{\phi_i\over\phi_0}\right)=
\pm\sin\psi_m\sin\left(\pi{\phi_i\over\phi_0}\right).
\end{equation}
We plug now Eq. (\ref{eqn_77}) into Eq. (\ref{eqn_72}) and obtain
\begin{equation}
\label{eqn_78}
\pm{2\pi\over\alpha}{\Lambda\over L}
{\phi_a-\phi_i\over\phi_0}=
\cos^2\left(\pi{\phi_i\over\phi_0}\right).
\end{equation}
\par
In the case of a long junction the LHS of Eq. (\ref{eqn_78}) is small.
As a result in the zero approximation in $\Lambda/L\ll 1$ the flux
inside the junction, $\phi_i$, is a constant defined by the roots of
equation $\cos(\pi\phi_i/\phi_0) =0$, {\it i.e.}, the values of
$\phi_i$ are given by $\phi_i=(n+1/2)\phi_0$, where $n= 0,\pm 1,\pm 2,
...$ is an integer. In the next approximation in $\Lambda/L\ll 1$ the
flux $\phi_i$ depends on the flux $\phi_a$ and we find
\begin{eqnarray}
\label{eqn_79}
\phi_i&=&\pm\sqrt{{2\over\pi\alpha}\,{\Lambda\over L}\,
{\tilde{\phi_a}\over\phi_0}}\,\,\phi_0 +
\left(n+{1\over 2}\right)\,\phi_0\,,
\\
\label{eqn_80}
\psi_m&=&\sqrt{{2\pi\over \alpha}\,{\Lambda\over L}\,
{\tilde{\phi_a}\over\phi_0}}\ll 1\,,
\end{eqnarray}
where
\begin{equation}
\label{eqn_81}
\tilde{\phi_a}=\phi_a-\left(n+{1\over 2}\right)\phi_0\,.
\end{equation}
\par
It follows therefore from the theoretical calculations that if the
applied field $H_a$ is increasing or decreasing, then inside the
intervals $(n-1/2)\phi_0<\phi_a<(n+1/2)\phi_0$ the flux in the bulk,
$\phi_i$, is almost constant. At the ends of these intervals the flux
$\phi_i$ ``jumps'' increasing or decreasing its value by one flux
quantum.
\par
Using Eq. (\ref{eqn_73}) we find that the maximum supercurrent in the
zero approximation in $\Lambda/L\ll 1$ is given by
\begin{equation}
\label{eqn_82}
I_m\approx 2\alpha I_c\,,
\end{equation}
{\it i.e.}, for long tunnel junctions ($L\gg\Lambda$) the value of
$I_m$ at high fields is almost field independent.
\par
\section{Numerical simulations}
\label{sec_IV}
\par
We used numerical simulations to calculate the maximum supercurrent in
a wide range of parameters characterizing Josephson tunnel junctions
with alternating critical current density. The computations were
performed by means of the time dependent sine-Gordon equation. The
spatially alternating critical current density were introduced by the
periodic function $g(x)$. In the dimensionless form this equation
yields
\begin{equation}
\label{eqn_83}
\ddot{\varphi}+\delta\dot{\varphi}-\varphi'' + [1+g(\zeta)]\sin\varphi
=0\,,
\end{equation}
where the dimensionless time $\tau =\Omega t$ and space $\zeta
=x/\Lambda$ variables are normalized by the Josephson frequency
$\Omega$ and length $\Lambda$, $\delta\ll 1$ is the damping constant,
\cite{Barone_1}
\begin{equation}
\label{eqn_84}
g(\zeta)=\sqrt{2\gamma}\,\,{2\pi\Lambda\over
l}\,\sin\left({2\pi\Lambda\over l}\,\zeta +\theta_0\right)\,,
\end{equation}
the phase shift $\theta_0$ defines the value of $\alpha$, $\alpha
=\sqrt{2\gamma}\,\cos\theta_0$, and $N=L/l$ is an integer ($N\gg 1$).
\par
The boundary conditions for Eq. (\ref{eqn_83}) are given by the set of
Eqs. (\ref{eqn_28}), (\ref{eqn_29}) and take the form
\begin{eqnarray}
\label{eqn_85}
\varphi_0'={2H_a\over H_s} + {I\over 2I_c},\\
\label{eqn_86}
\varphi_L'={2H_a\over H_s} - {I\over 2I_c}\,.
\end{eqnarray}
\par
The convergency criterion for solutions matching equations
(\ref{eqn_85}) and (\ref{eqn_86}) was based on the standard assumption
that after sufficiently large interval of time ($\tau\gg 1$) the
spatial average of $\dot{\varphi}^2(\zeta,\tau)$ fits the condition
$\langle\dot{\varphi}^2\rangle\le\delta_m^2$, where $\delta_m\ll 1$ is
a certain constant. We use a standard approach to calculate the maximum
value of the supercurrent $I_m$. Namely, for each value of the applied
field $H_a$ we find the current $I_m$ for which there is a solution of
Eq. (\ref{eqn_83}) matching boundary conditions (\ref{eqn_85}) and
(\ref{eqn_86}) and converging after a certain time $\tau_c\gg 1$, and
there is no solutions converging at $\tau\gg 1$ for currents higher
than $I_m$. We use the function $\varphi(\zeta,\tau_c)$ calculated for
the field $H_a$ as an initial condition $\varphi(\zeta,0)$ for the next
value of the field $H_a+\Delta H_a$, where $\Delta H_a\ll H_a$.
\par
\subsection{Finite difference scheme}
\label{sec_IV_a}
\par
We solved Eq. (\ref{eqn_84}) numerically using the leap frog method,
which was adopted to our case. We checked the stability and convergency
of the obtained solutions, and arrived at
\begin{eqnarray}
\label{eqn_87}
\varphi\ &\rightarrow&\
{\varphi^m_{\rm n-1}+\varphi^m_{\rm n+1}\over2} \equiv \tilde{\varphi}^m_n\,,\\
\label{eqn_88}
\dot{\varphi}\ &\rightarrow&\ {\tilde{\varphi}^m_n - \varphi^{\rm m-1}_n
\over\Delta_\tau}\,,\\
\label{eqn_89}
{\partial^2\varphi\over\partial\tau^2}\ &\rightarrow&\ {\varphi^{\rm
m+1}_n -2\varphi^m_n +\varphi^{\rm m-1}_n\over\Delta_\tau^2}\,,\\
\label{eqn_90}
{\partial^2\varphi\over\partial\zeta^2}\ &\rightarrow&\ {\varphi^m_{\rm
n+1}-2\varphi^m_n +\varphi^m_{\rm n-1}\over\Delta_
\zeta^2}\,,
\end{eqnarray}
where $\Delta_\tau$ and $\Delta_\zeta$ are steps along $\tau$ and
$\zeta$ axes correspondingly, the superscript $m$ is for the discrete
$\tau$ axis, and the subscript $n$ is for the discrete $\zeta$ axis.
Next, we choose $\Delta_\zeta$ to be equal to 1/12 of the period of the
rapidly alternating function $g(\zeta)$ and set $\Delta_\tau =
\Delta_\zeta$. As a result we arrive at the following final difference
scheme:
\begin{eqnarray}
\label{eqn_91}
\varphi^{\rm m+1}_n =-(1-\delta\Delta_\tau)\varphi^{\rm m-1}_n
+(2-\delta\Delta_\tau)\tilde{\varphi}^m_n\\
\nonumber
-\Delta_\tau^2\,(1+g_n)\sin\tilde{\varphi}^m_n.
\end{eqnarray}
\par
To obtain sufficiently accurate numerical data but to keep the time
which is necessary for the numerical simulations reasonable we choose
the convergency criterion and the value of the decay constant $\delta$
to be dependent on the length of the junction $L$. Specifically, we
used for convergency criterion the following relations
\begin{eqnarray}
\label{eqn_92}
\sqrt{\left\langle\dot{\varphi}^2\right\rangle}<10^{-7},\quad {\rm for}
\quad L\le 8\Lambda,\\
\label{eqn_93}
\sqrt{\left\langle\dot{\varphi}^2\right\rangle}<10^{-4},\quad {\rm for}
\quad L>8\Lambda\,.
\end{eqnarray}
The value of the decay constant $\delta$ of junctions with $L\le
8\Lambda$ was chosen from $\delta =2$ for $L=\Lambda/2$ to $\delta
=.25$ for $L=8\Lambda$. In the case of junctions longer than $8\Lambda$
we took $\delta$ to be dependent on the convergency rate
\begin{eqnarray}
\label{eqn_94}
\delta = 1.2, \quad {\rm if} \quad
\sqrt{\left\langle\dot{\varphi}^2\right\rangle}>10^{-7},\\
\label{eqn_95}
\delta =0.1, \quad {\rm if} \quad
\sqrt{\left\langle\dot{\varphi}^2\right\rangle} > 10^{-4}.
\end{eqnarray}
\par
\subsection{Results of numerical calculations}
\label{sec_IV_b}
\par
In this subsection we summarize the results of our numerical
simulations for short ($L\ll\Lambda$), long ($L\gg\Lambda$), and
intermediate ($L\sim\Lambda$) junctions and compare the numerically
calculated data to the theoretical results.
\par
\begin{figure}
\begin{displaymath}
\includegraphics[width=0.50\columnwidth]{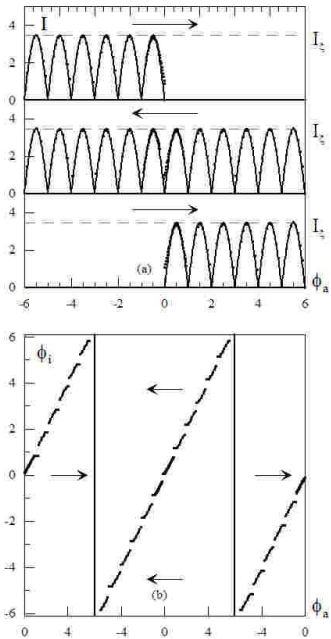}
\end{displaymath}
\caption{The maximum supercurrent $I_m$ and internal flux $\phi_i$
dependencies on the applied flux $\phi_a$ for a short junction
($L=0.25\,\Lambda$, $\alpha=2\sqrt{3}$ and $\gamma=6$). The arrows
indicate the sweeping direction of the applied flux, the points are for
the results of the numerical calculations, and the solid lines are for
the surface current $I_\zeta(\phi_a)$ given by Eq. (\ref{eqn_45}).}
\label{fig_3}
\end{figure}
%
In Fig. \ref{fig_3} (a) we demonstrate the dependence of the maximum
supercurrent on the applied flux, $I_m(\phi_a)$, for a short junction,
$L=0.25\,\Lambda$. In agreement with the theoretical results obtained
in Sec.~\ref{sec_III_a}~(see Eq.~(\ref{eqn_45})) we find that
$I_m(\phi_a)\approx I_{\zeta}(\phi_a)$ except for small deviations at
low fields. In Fig. \ref{fig_3} (b) we plot the internal flux $\phi_i$
as a function of the applied flux $\phi_a$. It is seen from the graphs
that $\phi_i\approx \phi_a$, which is in agreement with the assumptions
of the theoretical calculations of Sec. \ref{sec_III_a}. Small flux
``jumps'', $\Delta\phi\ll\phi_0$, are seen in Fig. \ref{fig_3} (b) in
the vicinity of $\phi_a=n\phi_0$, where $n$ is an integer. These small
flux jumps are generated by the high density screening currents
$\sim\langle |j_c(x)|\rangle\gg\langle j_c(x)\rangle$ flowing at the
edges of the junctions. The length of these current-carrying edges is
of the order of $l$ and therefore the value of $\Delta\phi$ can be
estimated as follows. First, using Maxwell equations we find the field
drop $\Delta H$ at the edges to be $\Delta H\approx
4\pi\langle|j_c|\rangle l/c$. Next, we estimate $\Delta\phi$ as a
product of the field drop $\Delta H$ and the effective area of the
junction, $2\lambda L$, {\it i.e.}, $\Delta\phi\approx 2\lambda L\Delta
H$. Finally, we write the parameter $\alpha$ as $\alpha\approx \langle
|j_c|\rangle\,l/\langle j_c\rangle\Lambda$. Combining these three
relations we find an estimate for $\Delta\phi$ in the form
\par
\begin{equation}
\label{eqn_96}
\Delta\phi\approx{\alpha\over2\pi}{L\over\Lambda}\phi_0.
\end{equation}
\par
It is worth noting that $\Delta\phi$ coincides with the coefficient in
Eq. (\ref{eqn_72}) for the difference between the internal flux and the
applied flux. Using the data $\alpha=2\sqrt{3}$ and $L=\Lambda/4$ we
obtain $\Delta\phi=0.14\phi_0$ which is in a good agreement with the
flux jumps shown in Fig. \ref{fig_3} (b).
\par
\begin{figure}
\begin{displaymath}
\includegraphics[width=0.50\columnwidth]{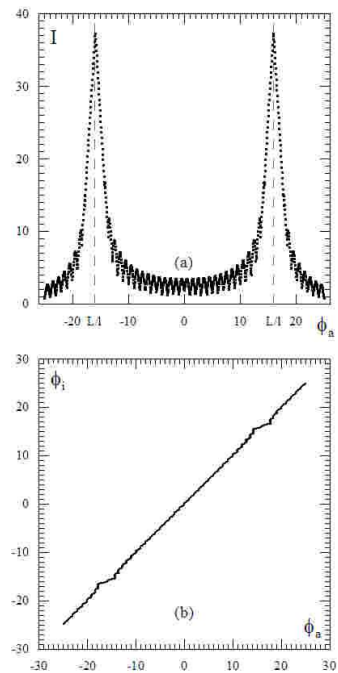}
\end{displaymath}
\caption{The maximum supercurrent $I_m$ and internal flux $\phi_i$
dependencies on the applied flux $\phi_a$ for a short junction
($L=0.5\,\Lambda$, $\alpha=2\sqrt{3}$ and $\gamma=6$). (a) The function
$I_m(\phi_a)$ with two side-picks located at $\phi_a=\pm
(L/l)\,\phi_0$; (b) the dependence $\phi_i(\phi_a)$ exhibiting
flux-plateaus at the side-picks.}
\label{fig_4}
\end{figure}
%
In this study we assume that the applied field is smaller than the
side-picks field $H_{\rm sp}$. The ``resonances'' at the side-picks are
discussed in detail in Ref. \onlinecite{Mints_1} and
\onlinecite{Buzdin_2}. We show in Fig. \ref{fig_4} the maximum
supercurrent $I_m(\phi_a)$ and internal flux $\phi_i(\phi_a)$ at the
side-picks for completeness and to reveal the flux-plateaus appearing
in the dependence $\phi_i(\phi_a)$ at $H_a=\pm H_{\rm sp}$.
\par
\begin{figure}
\includegraphics[width=0.50\columnwidth]{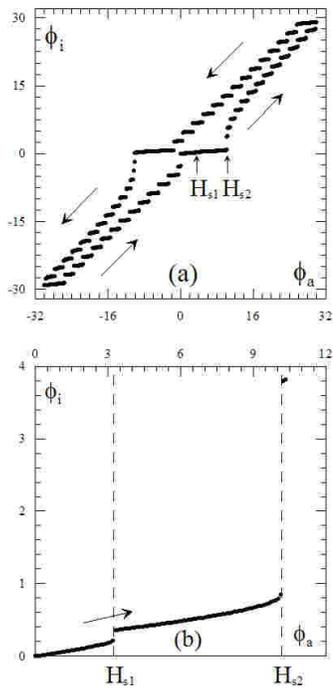}
\caption{The internal flux $\phi_i$ dependence on the applied flux
$\phi_a$ for a long junction ($L=30\,\Lambda$, $I=0$, $l=0.2\,\Lambda$,
$\alpha =0$ and $\gamma = 2$). The arrows indicate the direction of
sweeping of $\phi_a$ for the magnetization cycles starting from $\phi_a
=0$ and zero trapped flux. (a) The main features of internal flux
$\phi_i(\phi_a)$ of long junctions (flux-plateaus, flux jumps and
significant hysteresis); (b) the dependence $\phi_i(\phi_a)$ for the
applied field $H_a$ from the interval $0<H_a<H_{\rm s2}$.}
\label{fig_5}
\end{figure}
%
In Fig. \ref{fig_5} we show the internal flux $\phi_i$ for a long
junction, $L=30\Lambda$, as a function of the applied flux $\phi_a$.
The value of $\phi_i$ is less than one flux quanta if the field $H_a$
is lower than the second penetration field $H_{\rm s2}$. In this region
of fields the slope $d\phi_i/d\phi_a$ is proportional to $\Lambda/L\ll
1$, {\it i.e.}, it is almost zero. As a result, for long junctions in
low applied fields we observe two relatively long flux-plateaus. These
flux-plateaus, flux jumps and significant hysteresis in the
magnetization curves $\phi_i(\phi_a)$ are clearly seen in the whole
area of $\phi_a$. All these features of magnetization curves are in a
good agreement with the theoretical results obtained in Sec.
\ref{sec_III}.
\par
\begin{figure}
\begin{displaymath}
\includegraphics[width =0.60\columnwidth]{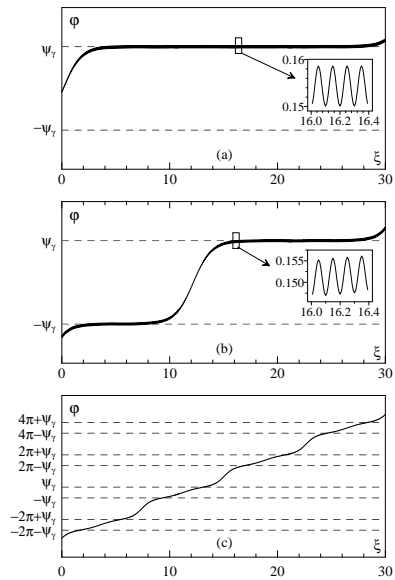}
\end{displaymath}
\caption{The phase $\varphi$ dependence on the coordinate $\zeta$ for a long
junction ($L=30\,\Lambda$, $I=0$, $l=0.2\,\Lambda$, $\alpha=0$,
$\gamma=2$). The insets show the oscillatory nature of the function
$\varphi(\zeta)$ on the space-scale of order $l$. (a) The applied field
is sweeping up, the applied flux $\phi_a =0.8\,\phi_0$; (b) the phase
$\varphi (\zeta)$ is shown after the first flux penetration that occurs
at $\phi_a=4.6\,\phi_0$, the applied field is sweeping up.}
\label{fig_6}
\end{figure}
%
In Figs. \ref{fig_6} (a), (b) we show the spatial distributions of the
phase $\varphi(\zeta)$ in a long junction, $L=30\,\Lambda$. The graph
in Fig. \ref{fig_6} (a) is obtained for a junction in the Meissner
state, {\it i.e.}, for the applied field $H_a$ from the interval
$0<H_a<H_{\rm s1}$. In this case the flux inside the junction is zero.
The graph shown in Fig. \ref{fig_6} (b) is calculated for a junction in
the one-splinter-vortex intermediate state, {\it i.e.}, for the applied
field from the interval $H_{\rm s1}<H_a<H_{\rm s2}$ and the internal
flux $\phi_i =\phi_1$ (see Eq. (\ref{eqn_58})). These numerical results
are in a good agreement with the theoretical calculation of Sec.
\ref{sec_III_b}.
\par
\begin{figure}
\begin{displaymath}
\includegraphics[width=0.50\columnwidth]{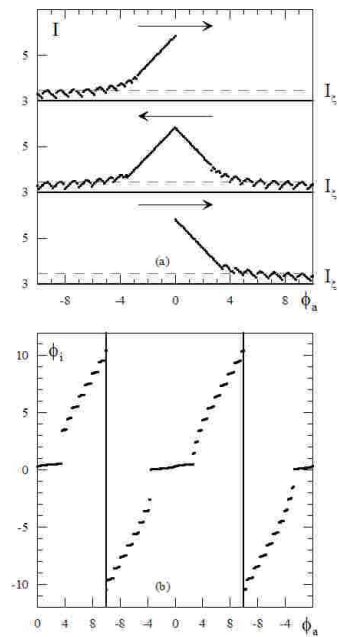}
\end{displaymath}
\caption{The maximum supercurrent $I_m$ and the internal flux $\phi_i$
dependencies on the applied flux $\phi_a$ for a long junction
($L=20\,\Lambda$, $\alpha=\sqrt{3}$, $\gamma=1.5$). The arrows indicate
the direction of sweeping of $\phi_a$ and the points are for the
results of the numerical calculations.}
\label{fig_7}
\end{figure}
%
In Fig. \ref{fig_7} (a) we plot the maximum value of the supercurrent
$I_m$ as a function of the applied flux $\phi_a$ for a long junction,
$L=20\,\Lambda$. At low applied flux the maximum current is linearly
dependent on $|H_a|$ yielding the middle triangle in agreement with Eq.
(\ref{eqn_69}). If the applied flux is sweeping up then the maximum
current in the mixed state is higher than the maximum current in the
Meissner state $H_a=H_{s2}-4\pi\alpha I_c/c$ and flux penetrates into
the junction (flux jump) and the dependence $I_m(\phi_a)$ changes. If
the applied field is sufficiently high then the maximum current is
approximately equal to $2\alpha I_c$ in agreement with Eq.
(\ref{eqn_82}). In Fig. \ref{fig_7} (b) we plot the internal flux
$\phi_i$ as a function of the applied flux $\phi_a$. As it is assumed
for fields lower than the first penetration field the junction is in
the Meissner state. When sweeping the field from low to high fields the
flux penetrates into the junction at $H_a=H_{s2}-4\pi\alpha I_c/c$
yielding a finite flux density. When sweeping the field from high to
low values the Josephson vortices leave the junction one by one
yielding the additional two steps between the plateau and the mixed
state. In the interval of high applied fields the flux jumps are of
order of one flux quanta $\phi_0$ and in between the flux is almost
constant in agreement with Eq. (\ref{eqn_79}).
\par
\begin{figure}
\begin{displaymath}
\includegraphics[width=0.50\columnwidth]{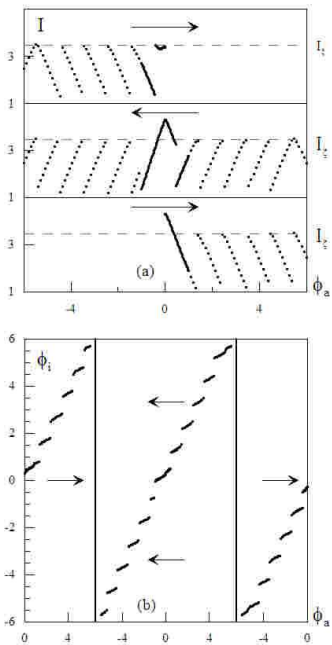}
\end{displaymath}
\caption{The maximum supercurrent $I_m$ and internal flux $\phi_i$
dependencies on the applied flux $\phi_a$ for a junction with an
intermediate length ($L=2\,\Lambda$, $\alpha=2\sqrt{3}$, $\gamma=6$).
The points are for the results of numerical simulations. The arrows
indicate the direction of sweeping of the applied flux.}
\label{fig_8}
\end{figure}
%
In Fig. \ref{fig_8} (a) we plot the maximum current as a function of
the applied flux for a junction with an intermediate length
$L=2\,\Lambda\sim\Lambda$. In Fig. \ref{fig_8} (b) we plot the flux
$\phi_i$ as a function of the applied flux. It is seen that the flux
$\phi_i$ differs from the applied flux $\phi_a$ by less then one flux
quantum $\phi_0$ as for the short junctions. The flux jumps happen at
$\phi_a=(n+1/2)\phi_0$, where $n$ is an integer. The value of
$\Delta\phi$ is well approximated by Eq. (\ref{eqn_96}).
\par
\section{Summary}
\label{sec_V}
\par
To summarize, we consider theoretically and numerically the maximum
supercurrent across Josephson tunnel junctions with a critical current
density, which is rapidly alternating along the junction. These complex
Josephson tunnel systems were treated recently in asymmetric grain
boundaries in thin films of high-temperature superconductor
YBa$_2$Cu$_3$O$_{7-x}$ and in superconductor-ferromagnet-superconductor
heterostructures.
\par
Our theoretical study is based on coarse-grained sine-Gordon equation.
We derive boundary conditions to this equation and find explicit
dependencies of the maximum supercurrent across a junction on the
magnetic field in the Meissner and mixed states for short and long
junctions. We show that in the case of a Josephson junction with
rapidly alternating critical current density there can exist
one-splinter-vortex mixed state and two flux-penetration fields. The
obtained theoretical results are verified by numerical simulations of
exact sine-Gordon equation. We demonstrate that the theoretical and
numerical results are in a good agreement.
\par
\begin{acknowledgments}
The authors are grateful to J. R. Clem, A. V. Gurevich, V. G. Kogan and
J. Mannhart for numerous stimulating discussions. CWS acknowledges the
support by the BMBF and by the DFG through the SFB 484.
\end{acknowledgments}
\par
\bibliography{Max_SCA}
\end{document}